\documentclass[amsmath,amssymb,amsbsy,prb,twocolumn,superscriptaddress]{revtex4-1}
\usepackage[dvipdfmx]{graphicx}
\usepackage[dvipdfmx]{color}
\usepackage{dcolumn}
\usepackage{bm}
\usepackage{braket}
\usepackage{mathtools}
\usepackage[dvipdfmx,breaklinks,colorlinks=true,linkcolor=blue,urlcolor=blue,citecolor=blue]{hyperref}

\begin{document}
\title{Tunable Spin Dynamics in Chiral Soliton Lattice}
\author{Kunio Tokushuku}
\affiliation{Department of Physics, University of Tokyo,
Hongo, Bunkyo-ku, Tokyo 113-0033, Japan }
\email{tokushuku@hosi.phys.s.u-tokyo.ac.jp}

\author{Jun-ichiro Kishinei}
\affiliation{Division of Natural and Environmental Sciences, The Open University of Japan, Chiba 261-8586, Japan}

\author{Masao Ogata}
\affiliation{Department of Physics, University of Tokyo,
Hongo, Bunkyo-ku, Tokyo 113-0033, Japan }

\begin{abstract}
We study the dynamics of a chiral soliton lattice (CSL) in a classical one-dimensional spin chain coupled to conduction electrons under an electric field. The CSL has attracted much interest because its period can be easily controlled by an external magnetic field. We clarify the dependence of the CSL dynamics on its period. A collective coordinate and an SU(2) gauge method are used for the analysis. It turns out that the velocity of the CSL decreases as the period becomes longer. We also mention the relation between the velocity and the magnetic resistance.
\end{abstract}

\maketitle

\section{Introduction}

\begin{figure}[]
\begin{center}
\includegraphics[width=\hsize]{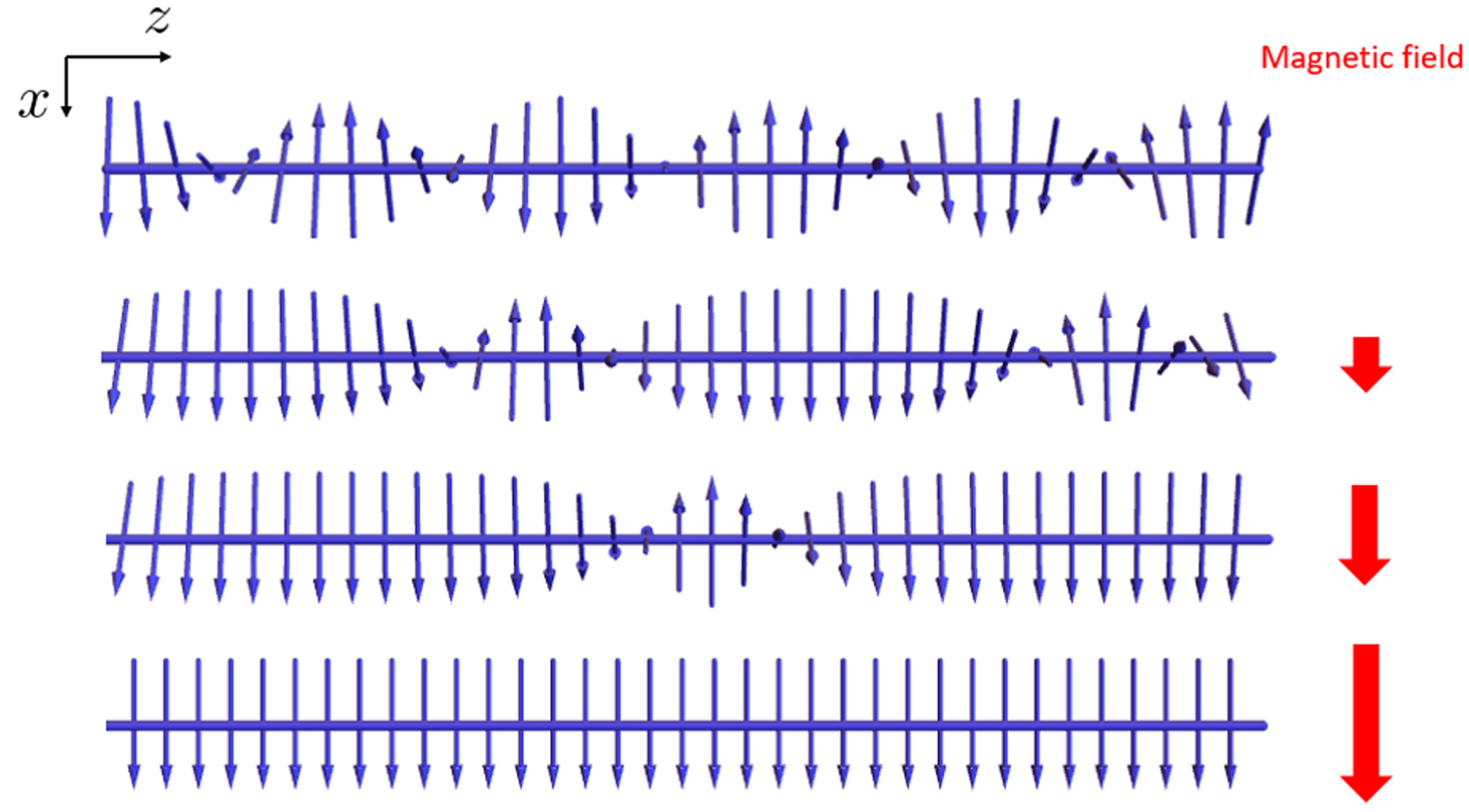}
\end{center}
\caption{Schematic picture of a chiral soliton lattice (CSL). The period of the CSL continually becomes longer from the CHM state to the FFM state.}
\label{fig:confirmation}
\end{figure}

Recently, the Dzyaloshinskii--Moriya (DM)\cite{Dzyaloshinsky1958,Moriya1960} interaction has attracted much attention. It is an interaction between spins that appears when the inversion symmetry is broken. One of the most interesting materials that have nonlinear spin structures is a chiral helical magnet (CHM); the CHM is a quasi-one-dimensional system with the DM interaction parallel to the spin axis. When an external magnetic field is applied perpendicular to the one-dimensional direction, there is competition between the Zeeman energy and the DM interaction. In this case, the period of the nonlinear spin structure continuously becomes longer when the magnetic field becomes larger; finally the system becomes a forced ferromagnet (FFM). Figure \ref{fig:confirmation} shows this situation. This tunable superlattice is called a chiral soliton lattice (CSL)\cite{Dzyaloshinsky1964,Izyumov1984,Kishine2005}. \par
Recently, several experiments have reported the realization of the CSL\cite{Zheludev1997,Togawa2012,Kanazawa2016} .
Inspired by these experiments, many experimental and theoretical studies have been carried out.
It turned out that many fascinating phenomena originate from the interaction between the CSL and conduction electrons.\cite{KishineReview}
Togawa et al.\cite{Togawa2013} revealed that the magnetic resistivity (MR) depends on the period of the CSL in $\mathrm{CrNb}_3\mathrm{S}_6$, which is one of the ideal materials realizing the CSL. They found a negative MR in a wide range of temperatures. The origin of this negative MR is ascribed to the decrease in soliton density, in accordance with the increasing period upon increasing the magnetic field strength.
In addition, discretization effects of the MR occur in a micrometer-size sample, where the number of solitons is limited to a several hundred\cite{Togawa2015}. \par
The above experiments show a fascinating feature of the CSL; one can easily control various responses of the CSL by an external magnetic field. It is also expected that the torque on the CSL induced by spin-polarized electric current can be controlled by an external magnetic field. This torque causes the dynamics of domain walls\cite{Berger1978,Yamaguchi2004,Tatara2004} and will have a large impact on its application\cite{Parkin2008,Hayashi2009}. However, the effects of changing the spin structure on the torques have not been well studied. There is a previous study on the dynamics of the CSL under an electrical current\cite{Kishine2010}, which shows that the CSL moves at a certain velocity after some relaxation time. However, in this work, only the limit of the weak magnetic field was considered and consequently the effects of a finite magnetic field on the torque were  left unaddressed. Therefore, the magnetic field dependence of the dynamics of the CSL is still unknown. In the present paper, we clarify the magnetic field dependence of the velocity of the CSL, starting from a microscopic model and using the SU(2) spin gauge transformation, which is applicable to an arbitrary spin structure. We will show that the torque from conduction electrons depends on the magnetic field and that the dynamics can be controlled.  \par
The outline of this paper is as follows. 
In section 2, we show how the CSL changes under an external magnetic field; we construct the CSL Lagrangian in section 3. In section 4, we solve the equation of motion. At this stage, we need to calculate the quantum expectation value of spins of the conduction electrons. Finally we show the period dependence of the CSL dynamics in section 5. We also mention the relation between the CSL dynamics and the MR.

\section{Model and the CSL}
We consider the CSL interacting with the conduction electrons through the s-d exchange interaction, which is described by the Hamiltonian, $\mathcal{H}= \mathcal{H}_{\mathrm{CSL}}+\mathcal{H}_{\mathrm{sd}}+\mathcal{H}_{\mathrm{el}}$ with
\begin{align}
\mathcal{H}_{\mathrm{CSL}} =&\int^{L(H^x)}_0 dz \frac{JS^2a_0}{2}\left[(\partial_z \theta)^2 +(\partial_z \varphi)^2\sin^2\theta\right] \nonumber \\
&-S^2 D(\partial_z\varphi)\sin^2\theta +\frac{S}{a_0}g\mu_B H^x \sin\theta\cos\varphi,  \label{eq:HCSL} \\
\mathcal{H}_{\mathrm{sd}}=&-\frac{ SJ_{\mathrm{sd}}}{a_0}\int d^3\bm x \bm n(z) \cdot \left(\hat{c}^\dagger(\bm x) \bm \sigma \hat{c}(\bm x)\right), \label{eq:Hsd} \\
\mathcal{H}_{\mathrm{el}}=&\int d^3\bm x \sum_{\sigma}\hat{c}_\sigma^\dagger(\bm x)\biggl[-\frac{\hbar^2}{2m_e}(\bm \nabla^2)-\mu_e\nonumber \\
&\:-\frac{e\hbar}{m_e}i\bm A_{\mathrm{em}}\cdot\bm \nabla\biggr] \hat{c}_{\sigma}(\bm x). \label{eq:Hel}
\end{align}
Here, we use the polar angle of the classical spin as $\bm S=S\bm n=S(\sin\theta\cos\varphi,\sin\theta\sin\varphi,\cos\theta)$, which gives the shape of the CSL. We will see later that $\theta$ and $\varphi$ represent the out-of-plane and in-plane angles, respectively. $J(>0)$ is the exchange interaction and $D$ is the coefficient of the DM interaction parallel to the one-dimensional direction. In the third term of Eq. (\ref{eq:HCSL}),
 $g$ is the gyromagnetic ratio, $\mu_B$ is the Bohr magneton, and $a_0$ is a lattice constant. Hereafter we treat $g\mu_B$, $a_0$, and $\hbar$ as 1 for simplicity.  $H^x$ is the external magnetic field acting on the local spins and is perpendicular to the one-dimensional direction (see Fig. \ref{fig:confirmation}).  It should be noted that $L(H^x)$ is a magnetic-field-dependent length that represents the period of the CSL; in the following, we sometimes write $L(H^x)$ as $L$.\par 
In Eqs. (\ref{eq:Hsd}) and (\ref{eq:Hel}), we denote the annihilation (creation) operators of electrons as $\hat{c}_{\sigma}$ ($\hat{c}_{\sigma}^{\dagger}$), where $\sigma=\pm$ represents the spin state. $J_{\mathrm{sd}}$ is the coefficient of the s-d interaction. We assume that the conduction electrons are three-dimensional, while the CSL does not depend on $x$ and $y$. In Eq. (\ref{eq:Hel}), we assume that the conduction electrons are not subjected to the external magnetic field, and $\bm A_{\mathrm{em}}$ is given by
\begin{align}
\bm A_{\mathrm{em}}=i\frac{\bm E}{\Omega_0}e^{i\Omega_0 t},
\end{align}
where $\bm E$ is the applied homogeneous electric field parallel to the one-dimensional direction. At the end of the calculation, the frequency $\Omega_0$ is set as $\Omega_0 \rightarrow 0$.

In this section, we analyze $\mathcal{H}_{\mathrm{CSL}}$. By minimizing the energy, we can see that the ground-state spin configuration is given by
\begin{align}
\sin\frac{\varphi_0(z)}{2}=&\mathrm{sn}\left(\frac{m}{\kappa}z,\kappa\right),  \label{eq:hamiltonian} \\
\theta(z)=&\frac{\pi}{2},
\end{align}
where $m^2=\frac{H^x}{JS}$, sn$(z,\kappa)$ is Jacobi's elliptic function\cite{Whittaker1927} and $\kappa$ $(0 \leq \kappa \leq 1)$ is the elliptic modulus (see Fig. \ref{fig:confirmation}). $\kappa$ is determined from the following relation:
\begin{align}
\frac{H^x}{H^x_c}=\left(\frac{\kappa}{E(\kappa)}\right)^2,
\end{align}
where $H^x_c \left[=\left(\frac{\pi D}{4J}\right)^2JS\right]$\cite{Kishine2012} is the critical magnetic field above which the system becomes an FFM. When the system is an FFM, $\kappa=1$. In addition, $\kappa=0$ and $0<\kappa<1$ correspond to the CHM and CSL, respectively. Typically, $H_c$ is not so high and $H^x_c \simeq 0.23T$ in the case of $\mathrm{CrNb}_3\mathrm{S}_6$ for instance\cite{Togawa2012}.
From the nature of the sn function, the period of the CSL is determined as
\begin{align}
L(H^x)=& \frac{8E(\kappa) K(\kappa)}{\pi q_0}, 
\end{align}
with $q_0=\frac{D}{J}$, and $K(\kappa)$ and $E(\kappa)$ are the complete elliptic integrals of the first kind and second kind, respectively.
The period $L(H^x)$ monotonically increases from $L(0)=\frac{2\pi}{q_0}$ to infinity. \par
Since we study the states under an electric field, it is necessary to know the excited states. Therefore, we introduce small deviations, $\delta\theta(z)$ and $\delta\varphi(z)$, of the local spins around the ground state as
\begin{align}
\theta(z)=&\frac{\pi}{2}+\delta\theta(z), \label{eq:yuragit}\\
\phi(z)=&\phi_0(z)+\delta\phi(z) .\label{eq:yuragip}
\end{align}
By substituting Eqs. (\ref{eq:yuragit}) and (\ref{eq:yuragip}) into Eq. (\ref{eq:HCSL}) and expanding the Hamiltonian with respect to $\delta\theta(z)$ and $\delta\varphi(z)$ up to the second order, we obtain 
\begin{align}
\partial\mathcal{H}_{\mathrm{CSL}}=\frac{JS^2a_0}{2}\int^L_0 dz (\delta\phi\hat{\Lambda}_\phi\delta\phi+\delta\theta\hat{\Lambda}_\theta\delta\theta) , \label{eq:deltaH}
\end{align}
where
\begin{align}
\hat{\Lambda}_\phi=&-\frac{m^2}{\kappa^2}(\partial^2_{\bar{z}}-2\kappa^2\mathrm{sn}^2\bar{z}+\kappa^2),  \\
\hat{\Lambda}_\theta=&\hat{\Lambda}_\phi+\Delta(z),  \\
\Delta(z)=&-(\partial_z\phi_0)^2+2q_0(\partial_z\phi_0),  
\end{align}
with $\bar{z}=(m/\kappa)z$. We set the ground-state energy of the CSL as 0. \par
To diagonalize Eq. (\ref{eq:deltaH}), we treat the inhomogeneous gap $\Delta(z)$ as its average value $\bar{\Delta}$,
\begin{align}
\bar{\Delta}=&\frac{1}{L}\int^L_0 dz\left[-(\partial_z\phi_0)^2+2q_0(\partial_z\phi_0)\right] \nonumber \\
=&\frac{\pi^2q_0^2}{4K(\kappa)E(\kappa)}.
\end{align}
This approximation is valid in the weak-field region. Then, the characteristics polynomials of Eq. (\ref{eq:deltaH}) are classified as the Lam$\acute{\mathrm{e}}$ equation\cite{Whittaker1927,Sutherland1973}.
We introduce eigenfunctions $\nu(z)$ and $u(z)$ and eigenvalues $\lambda^{(\theta)}$ and $\lambda^{(\varphi)}$ that satisfy the characteristics polynomials
\begin{align}
\partial_{\bar{z}}\nu(z)=&\left[2\kappa^2\mathrm{sn}^2\bar{z}-\kappa^2-\left(\frac{\kappa}{m}\right)^2\lambda^{(\varphi)}\right]\nu(z),  \\
\partial_{\bar{z}}u(z)=&\left[2\kappa^2\mathrm{sn}^2\bar{z}-\kappa^2-\left(\frac{\kappa}{m}\right)^2(\lambda^{(\theta)}-\Delta_0)\right]u(z) .
\end{align}
They are labeled by an index $q$ and given by
\begin{align}
\nu_q(z)=u_q(z)=N\frac{\mathcal{\theta}_4\left(\frac{\pi}{2K(\kappa)}\left(\bar{z}-\xi_q\right)\right)}{\mathcal{\theta}_4\left(\frac{\pi}{2K(\kappa)}\bar{z}\right)}e^{-i\frac{\kappa}{m}q\bar{z}} ,
\end{align}
where $\mathcal{\theta}_4(z)$ is Jacobi's theta function, $N$ is a normalization constant, and $\xi_q$ is a shift parameter\cite{Sutherland1973,Kishine2012}.
$\nu_q$ and $u_q$ are orthonormal eigenstates satisfying\par
\begin{align}
\int^L_0dz\nu_q(z)\nu_{q'}(z)=&\delta_{q,q'}, \\
\int^L_0dzu_q(z)u_{q'}(z)=&\delta_{q,q'} .
\end{align}\par
\begin{figure}[]
\begin{center}
\includegraphics[width=\hsize]{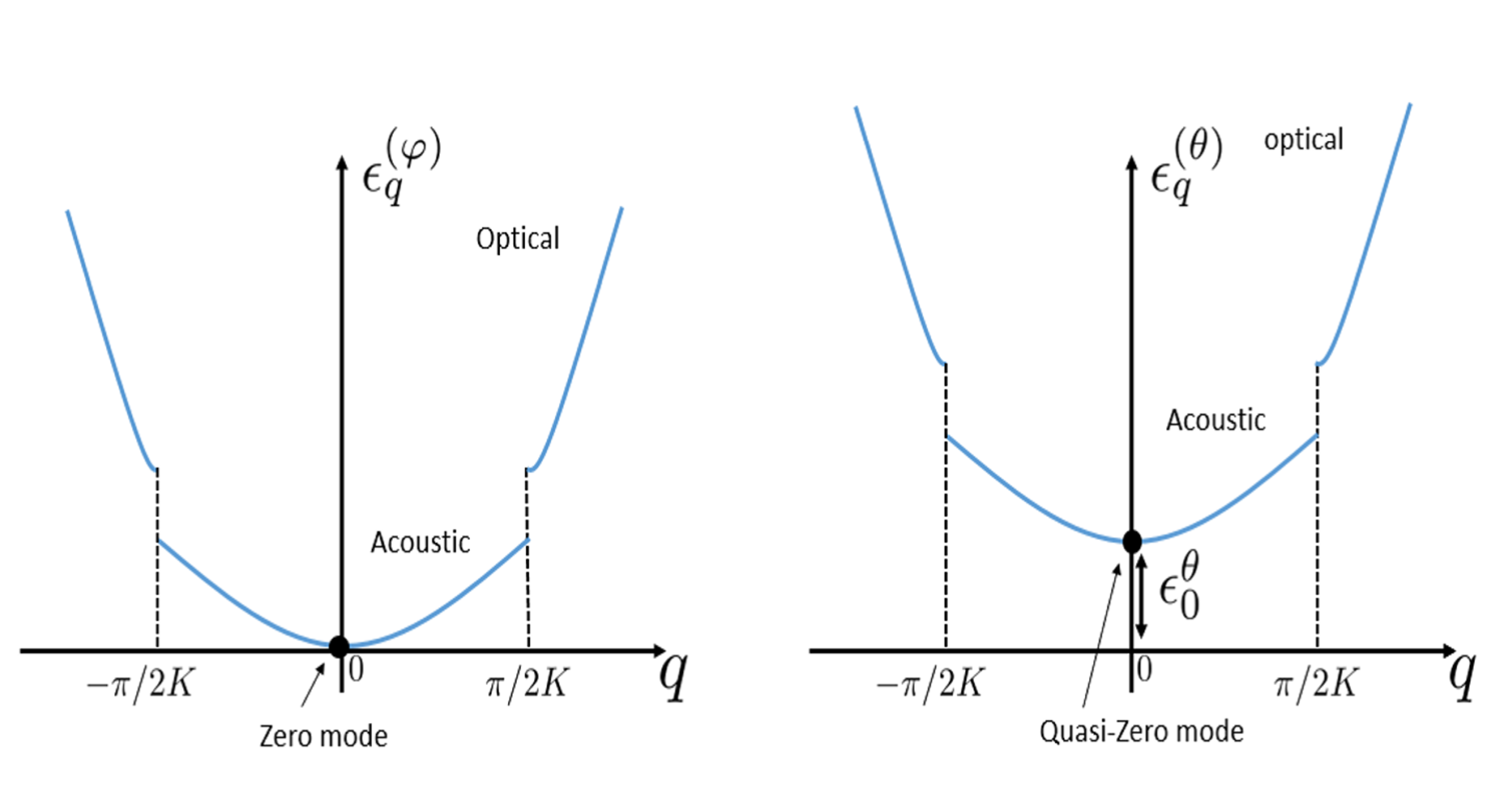}
\end{center}
\caption{Spectra of CSL: the in-plane mode $\epsilon^{(\varphi)}_q$ is gapless while the out-of-plane mode $\epsilon^{(\theta)}_q$ has a gap. There are other gaps at $\pm\pi/2K$, where 
a junction between an acoustic branch and an optical branch exists\cite{Sutherland1973}.
}
\label{fig:dispersion}
\end{figure}

Using these orthonormal bases, we expand the polar coordinates $\varphi(z,t)$ and $\theta(z,t)$ as
\begin{align}
\delta\varphi(z,t)&=\sum_{q}\eta_q(t)\nu_q(z),  \\
\delta\theta(z,t)&=\sum_{q}\xi_q(t)u_q(z) ,
\end{align}
where $\eta_q(t)$ and $\xi_q(t)$ are the coefficients of each mode.
Then, $\delta H_{\mathrm{CSL}}$ is diagonalized as
\begin{align}
\partial\mathcal{H}_{\mathrm{CSL}}=\sum_q[\epsilon_q^{(\varphi)}\eta_q^2(t)+\epsilon_q^{(\theta)}\xi_q^2(t)]. \label{eq:deriHCSL}
\end{align}
In the weak-field approximation, $\epsilon_q^{(\theta)}$ has the simple relation\cite{Kishine2012} 
\begin{align}
\epsilon_q^{(\theta)}=\epsilon_q^{(\varphi)} + \frac{JS^2a_0}{2}\Delta_0 .
\end{align}
The spectra of the CSL are shown in Fig. \ref{fig:dispersion}. The in-plane mode $\epsilon_q^{(\varphi)}$ is gapless; it is related to the renormalized Klein-Gordon equation\cite{Sutherland1973}. On the other hand, the out-of-plane mode $\epsilon_q^{(\theta)}$ has a finite gap, which corresponds to the energy tilting out of the easy plane caused by the DM interaction. The bottoms of the dispersions are called the zero mode and quasi-zero mode, respectively.

\section{Lagrangian and Equation of Motion}
First, we construct the Lagrangian. Hereafter, we only consider the quasi-zero mode and ignore the other modes. This approximation corresponds to the assumption that the CSL is sufficiently rigid. In this case, the polar coordinates are written as 
\begin{align}
\varphi(z,t)&=\varphi_0\left(z-Z(t)\right), \\
\theta(z,t)&=\frac{\pi}{2}+\xi_0(t)u_0\left(z-Z(t)\right),
\end{align}
where $Z(t)$ represents the coordinate of the CSL.
Since the quasi-zero mode $u_0$ has a finite energy $\epsilon_0^{(\theta)}$ while the $\varphi$-mode is gapless, Eq. (\ref{eq:deriHCSL}) is now given by
\begin{align}
\partial\mathcal{H}_{\mathrm{CSL}}=\epsilon_0^{(\theta)}\xi_0^2(t).
\end{align}
As a result, the only dynamical variables are $\xi_0(t)$ and $Z(t)$. Thus, we reconstruct the Lagrangian with these two variables. The Lagrangian of the spin system is written as
\begin{align}
\mathcal{L}=\mathcal{L}_{\mathrm{Berry}}-\partial\mathcal{H}_{\mathrm{CSL}}-\mathcal{H}_{\mathrm{sd}}, \label{eq:lagrangian}
\end{align}
where 
\begin{align}
\mathcal{L}_{\mathrm{Berry}}=S\int^L_0dz(\cos\theta-1)\partial_t\varphi, \label{eq:berryphase}
\end{align}
and
\begin{align}
\partial\mathcal{H}_{\mathrm{CSL}}&=\epsilon_0^{(\theta)}\xi_0^2(t), \\
\mathcal{H}_{\mathrm{sd}}&=-SJ_{\mathrm{sd}}\int^L_0dz <\hat{\bm s}(z,t)>\cdot\bm n(z,t).
\end{align}
Here, we treat spins of the conduction electrons as their expectation values in $\mathcal{H}_{\mathrm{sd}}$. Equation (\ref{eq:berryphase}) is the Berry phase term, which describes the time development of spins.
Since there are only two variables $\xi_0(t)$ and $Z(t)$, Eq. (\ref{eq:berryphase}) can be written as
\begin{align}
\mathcal{L}_{\mathrm{Berry}}&=S\int^L_0dz(\cos\theta-1)\partial_t \varphi \nonumber \\
&=S\int^L_0dz\left(\partial_x\varphi_0u_0\left(z-Z(t)\right)\xi_0(t)\right)\dot{Z}(t) \nonumber \\
&=S\mathcal{K}\xi_0(t)\dot{Z}(t), 
\end{align}
where
\begin{align}
\mathcal{K}=&\int^L_0dzu_0(z-Z(t))\partial_z\varphi_0 \nonumber \\
=&\sqrt{2\pi q_0}.
\end{align}
Note that $\mathcal{K}$ is independent of the magnetic field.\par
To take into account a damping effect, we include the Rayleigh dissipation term 
\begin{align}
\mathcal{W}_{\mathrm{Rayleigh}}&=\frac{\alpha S}{2}\int^L_0dz(\partial_t\theta)^2+\sin^2\theta(\partial_t\varphi)^2 \nonumber \\
&=\frac{\alpha S}{2}\int^L_0dz\left(\dot{\xi}_0^2(t)u_0^2+\sin^2\theta(\partial_z\varphi)^2\dot{Z}^2(t) \right) \nonumber \\
&=\frac{\alpha S}{2}\left(\mathcal{M}\dot{Z}^2(t)+\dot{\xi}^2_0(t)\right) ,
\end{align}
where 
\begin{align}
\mathcal{M}=\int^L_0dz(\partial_z\varphi_0)^2=2\pi q_0,
\end{align} 
and $\alpha$ is a small coefficient $\alpha =$0.01--0.1, representing the Gilbert damping\cite{Gilbert2004}.\par
Using the Euler-Lagrange-Rayleigh equation\cite{Goldstein2001}, the equation of motion becomes
\begin{align}
-S\mathcal{K}\dot{Z}(t)+2\epsilon_0^{(\theta)}\xi_0(t)-\tau(<\hat{\bm s}(z,t)>)&=-\alpha S\dot{\xi}_0(t) ,\label{eq:eom2} \\
S\mathcal{K}\dot{\xi}_0(t)-F(<\hat{\bm s}(z,t)>)&=-\alpha S\mathcal{M}\dot{Z}(t). \label{eq:eom3}
\end{align}
Hereafter, we only take $\xi_0(t)$ and $Z(t)$ up to linear terms in the equation of motion. $F(<\hat{\bm s}(z,t)>)$ and $\tau(<\hat{\bm s}(z,t)>)$ are the force and torque to the local spins due to the conduction electrons, respectively, which are expressed as
\begin{align}
F(<\hat{\bm s}(z,t)>)\equiv& J_{\mathrm{sd}}S\frac{\int^L_0 dz\delta \bm n[Z,\xi_0]\cdot<\hat{\bm s}(z,t)>}{\delta Z} \nonumber \\ 
=&-J_{\mathrm{sd}}S\int^L_0 dz(\partial_z\bm n)\cdot<\hat{\bm s}(z,t)> \nonumber \\
=&-J_{\mathrm{sd}}S\int^L_0 dz(\partial_z\theta)<\hat{s}_\theta(z,t)> \nonumber  \\
&+\sin\theta(\partial_z\varphi)<\hat{s}_\varphi(z,t)>,  \label{eq:force} \\ 
\tau(<\hat{\bm s}(z,t)>) \equiv&J_{\mathrm{sd}}S\frac{\int^L_0 dz\delta \bm n[Z,\xi_0]\cdot<\hat{\bm s}(z,t)>}{\delta \xi_0} \nonumber \\
=&J_{\mathrm{sd}}S\int^L_0 dzu_0\left(z-Z(t)\right)\nonumber \\
&\times(\bm n \times <\hat{\bm s}(z,t)>)\cdot\bm e_\varphi  \nonumber \\
=&J_{\mathrm{sd}}S\int^L_0 dzu_0\left(z-Z(t)\right) \nonumber\\
&\times\sin\theta <\hat{s}_\theta(z,t)>,\label{eq:tau}
\end{align}
with 
\begin{align}
<\hat{s}_\theta(z,t)>\equiv <\hat{\bm s}(z,t)>\cdot \bm e_\theta , \\
<\hat{s}_\varphi(z,t)>\equiv<\hat{\bm s}(z,t)>\cdot \bm e_\varphi,
\end{align}
where $\bm e_\theta(z,t)$ and $\bm e_\varphi(z,t)$ are the unit vectors in the $\theta$ and $\varphi$ directions in the spin space. Note that the $\theta$ and $\varphi$ directions depend on $z$ and $t$, and we only need the components of $<\hat{\bm s}(z,t)>$ perpendicular to $\bm n(z,t)$.

\section{Motion of the CSL under an Electric Field}
In this section, we analyze the dynamics of the CSL by solving Eqs. (\ref{eq:eom2})-(\ref{eq:tau}).
So far, Eqs. (\ref{eq:force}) and (\ref{eq:tau}) are exact. To calculate these terms, we need to analyze the quantum expectation value $<\hat{\bm s}(z,t)>$, which is obtained from the Hamiltonian of the conduction electrons, $\mathcal{H}_{\mathrm{el}}+\mathcal{H}_{\mathrm{sd}} $ [Eqs. (\ref{eq:Hsd}) and (\ref{eq:Hel})]. \par
Since we change the period of the CSL, we cannot use the hopping gauge method used in the preceding work\cite{Kishine2010}, which is valid only in the CHM case. Instead, in this paper, we use a local gauge transformation in spin space that diagonalizes the s-d coupling such that
\begin{align}
U^\dagger(z ,t)\left(\bm n\cdot\bm \sigma\right)U(z,t)=\sigma_z,
\end{align}
where $\bm \sigma$ is the Pauli matrix. $U(z,t)$ is a 2$\times$2 unitary matrix and its explicit form is given by
\begin{align}
U(z,t)&=\bm m(z,t)\cdot \bm \sigma ,  \\
\bm m(z,t) &=\left(\sin \frac{\theta}{2}\cos \varphi ,\sin\frac{\theta}{2}\sin\varphi,\cos\frac{\theta}{2}\right).
\end{align}
After this unitary transformation, a new electron operator $\hat{a}(\bm x,t)$ is defined as 
\begin{align}
\hat{c}(\bm x,t)=U(z,t)\hat{a}(\bm x,t).
\end{align}
In this framework, $\mathcal{H}_{\mathrm{sd}}$ in Eq. (\ref{eq:Hsd}) becomes $\tilde{\mathcal{H}}_{\mathrm{sd}}=-SJ_{\mathrm{sd}}\int d^3\bm x \hat{a}^\dagger(\bm x)\sigma_z\hat{a}(\bm x)$. Alternatively, the electrons represented by $\hat{a}(\bm x,t)$ are subjected to the SU(2) gauge field, which arises from $\partial_\mu \hat{c}(\bm x,t) = U(z,t)(\partial_\mu+iA_\mu(z,t))\hat{a}(\bm x,t)$, where $\mu=0$ or $z$ ($\partial_0=\partial/\partial t$) with
\begin{align}
A_\mu(z,t) &=(-i)U^{-1}(z,t)\partial_\mu U(z,t) \nonumber \\
&=(\bm m\times \partial_\mu \bm m)\cdot\bm \sigma \nonumber \\
&=\bm A_\mu(z,t)\cdot\bm\sigma ,
\end{align}
and
\begin{align}
\bm A_\mu(z,t)=\frac{1}{2}\left(\begin{array}{c}
-\partial_\mu\theta\sin\varphi-\sin\theta\cos\varphi\partial_\mu\varphi \\
\partial_\mu\theta\cos\varphi-\sin\theta\sin\varphi\partial_\mu\varphi \\
(1-\cos\theta)\partial_\mu\varphi \\ \end{array}
\right). \label{eq:formofA}
\end{align}
Note that only $\bm A_z(z,t)$ and $\bm A_0(z,t)$ are nonzero.\par
By this gauge transformation, the Lagrangian for the conduction electrons ($\mathcal{L}_{\mathrm{electron}}=\int d^3\bm x(i\hbar c^\dagger \dot{c})-\mathcal{H}_{\mathrm{el}}-\mathcal{H}_{\mathrm{sd}}$) becomes

\begin{align}
\mathcal{L}_{\mathrm{electron}}&=\sum_{\bm k,\sigma}i\hbar\hat{a}_{\bm k\sigma}^\dagger(\partial_t-\epsilon_{\bm k ,\sigma})\hat{a}_{\bm k\sigma}-\mathcal{H}_A,
\end{align}
\begin{align}
\mathcal{H}_A=&\sum_{\bm k \bm q} \biggl[ \frac{1}{m_e}\left(k_z+\frac{q}{2}\right)\bm A_z (-q)\hat{a}_{\bm k+\bm q}^\dagger \bm\sigma \hat{a}_{\bm k}  \nonumber\\
&\: +\bm A_0(-q)\hat{a}_{\bm k+\bm q}^\dagger \bm\sigma \hat{a}_{\bm k} \biggr] +\frac{ie E}{m_e\Omega_0}e^{i\Omega_0 t}\biggl[\sum_{\bm k}k_z\hat{a}_{\bm k}^\dagger \hat{a}_{\bm k} \nonumber \\
&+\sum_{\bm k \bm q}\bm A_z(-q)\hat{a}_{\bm k+\bm q}^\dagger \bm \sigma \hat{a}_{\bm k}\biggr] +O(A_z^2) ,
\end{align}
where $\bm q=(0,0,q)$,  $\bm A_\mu(q,t)\equiv \int^{L}_0 dz \bm A_\mu(z,t) e^{iq z} $ represents the Fourier transform of $\bm A_\mu(z,t)$, and
\begin{align}
\epsilon_{\bm k ,\pm}=&\frac{k^2}{2m_e}-\mu_e\mp J_{\mathrm{sd}}S . \label{eq:epsilonkpm}
\end{align}
$q$ represents the momentum of the CSL. Hereafter, we treat $\bm A_\mu$ perturbatively because the structure of the CSL changes slowly in a real space.\par
To study the dynamics of the CSL, we need to calculate $<\hat{\bm s}(z,t)>$ for a conduction electron. The expectation value $<\hat{\bm s}(z,t)>$ is obtained as
\begin{align}
<\hat{\bm s}(z,t)>=-i\mathrm{Tr}[G^<_{z\sigma , z\sigma'}(t,t)\bm\sigma_{\sigma , \sigma'}], \label{eq:cals}
\end{align}
with
\begin{align}
G^<_{z\sigma,z'\sigma'}(t,t')\equiv& i<\hat{c}_{\sigma'}^\dagger(z',t')\hat{c}_\sigma(z,t)> \nonumber \\
=&<\hat{a}_{\tau'}^\dagger(z',t')U^\dagger_{\tau',\sigma'}(z',t')U_{\sigma \tau}(z,t)\hat{a}_\tau(z,t)>,
\end{align}
which is the Keldysh lesser Green function. We define spin density without the factor $\frac{1}{2}$. Tatara et al.\cite{Tatara2007} obtained this quantity for general cases using diagrammatic perturbation theory at $T=0$. Using their results, we obtain
\begin{align}
<\hat{s}_\theta(z,t)>&=<\hat{s}_\theta^{(0)}(z,t)>+<\hat{s}_\theta^{(1)}(z,t)>, \label{eq:sthetato} \\
<\hat{s}_\varphi(z,t)>&=<\hat{s}_\varphi^{(0)}(z,t)>+<\hat{s}_\varphi^{(1)}(z,t)>, \label{eq:svarphito}
\end{align}
where
\begin{subequations}
\begin{align}
<\hat{s}_\theta^{(0)}(z,t)>&=\frac{-1}{J_{\mathrm{sd}}S}\sum_{q}e^{-iq z}\biggl[\left(\left(\bm e_\varphi\times\bm e_z\right)\cdot\bm A_0\left(q,t\right)\right)  \nonumber \\
&\times \chi_1^{(0)}(q) + \left(\bm e_\varphi\cdot\bm A_0\left(q,t\right)\right)\chi_2^{(0)}\left(q\right)\biggr] , \\ \label{eq:stheta0}
<\hat{s}_\varphi^{(0)}(z,t)>&=\frac{-1}{J_{\mathrm{sd}}S}\sum_{q}e^{-iq z}\biggl[\left(\bm e_\varphi\cdot\bm A_0\left(q,t\right)\right)\chi_1^{(0)}(q) \nonumber \\
& - \left(\left(\bm e_\varphi\times\bm e_z\right)\cdot\bm A_0\left(q,t\right)\right)\chi_2^{(0)}\left(q\right)\biggr] , \\
<\hat{s}_\theta^{(1)}(z,t)>&=\frac{-E}{J_{\mathrm{sd}}S}\sum_{q}e^{-iq z}\biggl[\left(\left(\bm e_\varphi\times\bm e_z\right)\cdot\bm A_z\left(q,t\right)\right)\nonumber \\
&\times \chi_1^{(1)}\left(q\right)+ \left(\bm e_\varphi\cdot\bm A_z\left(q,t\right)\right)\chi_2^{(1)}\left(q\right)\biggr] , \\
<\hat{s}_\varphi^{(1)}(z,t)>&=\frac{-E}{J_{\mathrm{sd}}S}\sum_{q}e^{-iq z}\biggl[\left(\bm e_\varphi\cdot\bm A_z\left(q,t\right)\right)\chi_1^{(1)}(q) \nonumber \\
&- \left(\left(\bm e_\varphi\times\bm e_z\right)\cdot\bm A_z\left(q,t\right)\right)\chi_2^{(1)}\left(q\right)\biggr], \label{eq:svarphi1}
\end{align}
\end{subequations}
with $\bm e_z$ being (0,0,1) and
\begin{subequations}
\begin{align}
\chi_1^{(0)}(q)&=\frac{2J_{sd}S}{V}\sum_{\bm k,\pm}\frac{f_{\bm k \pm}}{\epsilon_{\bm k+\bm q}-\epsilon_{\bm k}\pm2J_{sd}S} , \\
\chi_2^{(0)}(q)&=\frac{2 J_{sd}S}{V}\sum_{\bm k,\pm}\frac{\pi}{2}(f_{\bm k+}-f_{\bm k-})\delta\left(\epsilon_{\bm k+\bm q}-\epsilon_{\bm k}\pm2J_{sd}S\right) , \\
\chi_1^{(1)}(q)&=\frac{e\tau J_{sd}S}{3\pi m_e^2V}\sum_{\bm k \pm}\left(\bm k\cdot(\bm k+\frac{\bm q}{2})\right)i\frac{g^r_{\bm k\pm}-g^a_{\bm k \pm}}{\epsilon_{\bm k+\bm q}-\epsilon_{\bm k}\pm2J_{sd}S} , \\
\chi_2^{(1)}(q)&=\frac{e\tau J_{sd}S}{3\pi m_e^2V}\sum_{\bm k \pm}\left(\pm\frac{\pi}{2}\right)\left(\bm q\cdot\left(\bm k+\frac{\bm q}{2}\right)\right) \nonumber\\
&\times \delta(\epsilon_{\bm k+\bm q}-\epsilon_{\bm k}\pm2J_{sd}S)i(g^r_{\bm k\pm}-g^a_{\bm k \pm}). \label{eq:lastt} 
\end{align}
\end{subequations}
Here, $<\hat{s}_\theta^{(1)}(z)>$ and $<\hat{s}_\varphi^{(1)}(z)>$ are the terms proportional to $E$ and the $\Omega_0\rightarrow 0$ limit has been taken. In $\chi^{(0)}_1(q)$ and $\chi^{(0)}_2(q)$, $f_{\bm k\pm}=\Theta(-\epsilon_{\bm k,\pm})$ is the Fermi distribution function, where $\Theta(x)$ represents the step function. In $\chi^{(1)}_1(q)$ and $\chi^{(1)}_2(q)$, $g^{r(a)}_{\bm k \pm}=(-\epsilon_{\bm k,\pm}\pm\frac{i}{2\tau})^{-1}$. Here the relaxation time $\tau$ is phenomenologically introduced. \par
As a further approximation, we consider an adiabatic approximation\cite{Tatara2008}. In this approximation, we assume that the spin of a conduction electron completely follows the local spin $S\bm n(z,t)$. This situation is realized by neglecting the $q$ dependences of $\chi_j^{(i)}(q)$ ($i=0,1, j=1,2$), i.e., by approximating $\chi_j^{(i)}(q)$ as $\chi_j^{(i)}(0)$. In the present case, we obtain 
\begin{subequations}
\begin{align}
\chi_1^{(0)}(0)&=s, \\
\chi_2^{(0)}(0)&=0, \\
\chi_1^{(1)}(0)&=\frac{j_{\rm{s}}}{E}, \\
\chi_2^{(1)}(0)&=0, 
\end{align}
\end{subequations}
where
\begin{align}
s&=n_+ -n_- \\
j_s&=\frac{e^2\tau}{m}(n_+ -n_-) E,
\end{align}
are the spin density and spin current density of conduction electrons, respectively, and $n_\pm$ is the density of electrons with spin $\pm$ in the $\hat{a}_{\bm k}$ framework determined from $\epsilon_{\bm k ,\pm}$ in Eq. (\ref{eq:epsilonkpm}). Substituting these values into Eqs. (\ref{eq:sthetato}) and (\ref{eq:svarphito}), we obtain\cite{Tatara2008}
\begin{align}
<\hat{s}_\theta^{(ad)}(z,t)>&=-\frac{1}{SJ_{\mathrm{sd}}}[sA_0^\theta(z,t)+\frac{j_s}{e}A^\theta_z(z,t)] , \label{eq:adiabatic1} \\ 
<\hat{s}_\varphi^{(ad)}(z,t)>&=-\frac{1}{SJ_{\mathrm{sd}}}[sA_0^\varphi(z,t)+\frac{j_s}{e}A^\varphi_z(z,t)] , \label{eq:adiabatic2}
\end{align}
where $A^\theta_\mu\equiv\bm e_\theta\cdot\bm A_\mu$ and $A^\varphi_\mu\equiv\bm e_\theta\cdot\bm A_\mu$. From Eq. (\ref{eq:formofA}), $A^\theta_\mu$ and $A^\varphi_\mu$ become
\begin{subequations}
\begin{align}
A^\theta_0&=\frac{1}{2}\dot{Z}(t)(\partial_z\varphi), \\
A_z^\theta&=-\frac{1}{2}(\partial_z\varphi), \\
A_0^\varphi&=\frac{1}{2}u_0\left(z-Z(t)\right)\dot{\xi}(t) , \\
A_z^\varphi&=\frac{1}{2}\partial_z u_0\left(z-Z(t)\right)\xi(t).
\end{align}
\end{subequations}
After some algebra, Eqs. (\ref{eq:force}) and (\ref{eq:tau}) become
\begin{align}
F^{(\mathrm{ad})}&=-\frac{s\mathcal{K}}{2}\dot{\xi}_0(t) , \\ 
\tau^{(\mathrm{ad})}&=\frac{\mathcal{K}}{2}\left[s\dot{Z}(t)-\frac{j_s}{e}\right]. \label{eq:adtorque} 
\end{align}\par
The above results are in the adiabatic approximation. However, in this approximation, the dynamics of the CSL stops after a certain relaxation time, which is unphysical. Therefore, we consider the nonadiabatic force $F^{(\mathrm{non-ad})}$ in addition to $F^{(\mathrm{ad})}$. In the adiabatic approximation, the expectation value for conduction electrons depends only on the spin configuration $\bm n(z,t)$ at the same position. However, the nonlocal contributions with $q\neq 0$ in $\chi_j^{(i)}(q)$ neglected in the adiabatic approximation give the nonadiabatic force. In the present case, $F^{(\mathrm{non-ad})}$ acts in the same way as the so-called $\beta$ term\cite{Zhang2004,Thiaville2005}. \par
Using a similar method to that of Tatara et al.\cite{Tatara2007}, we obtain
\begin{align}
F^{(\mathrm{non-ad})}=&\sum_{\bm k,q,\pm}\frac{4\pi eE\tau S^2J_{\mathrm{sd}}^2}{mV}\delta(\epsilon_{\bm k\pm})\delta (\epsilon_{\bm k +\bm q\mp}-\epsilon_{\bm k\pm}) \nonumber \\
&\times A^\pm_z(q,t)A^\mp_z(-q,t) \nonumber \\
=&\frac{S^2J_{\mathrm{sd}}^2}{\pi}\left(\frac{ne\tau E}{m_e}\right)\sum_{q}\frac{f(q,t)}{4q}\nonumber\\
&\times\Theta\left(1+\frac{q}{2k_F}\right)\Theta\left(-\frac{q}{2k_F}+1\right) , \label{eq:owari} 
\end{align} 
with
\begin{align}                   
f(q,t)=&\frac{q}{2}\left[|\int^L_0dz\cos\varphi e^{iqz}|^2+|q\int^L_0dz\sin\varphi e^{iqz}|^2\right]. \label{eq:smallfq} 
\end{align} 
In calculating $F^{(\mathrm{non-ad})}$, we used $\theta=\frac{\pi}{2}$ and $\varphi = \varphi_0$ for simplicity. In other words, we have ignored the effect of the quasi-zero mode.
 We have also assumed that $\epsilon_{\rm{F}}>>SJ_{\rm{sd}}$. We will see later that $F^{(\mathrm{non-ad})}$ represents the reflection of the conduction electrons by the CSL.\par
 Nonadiabatic torque contributions, which arise in the same manner as $F^{(\mathrm{non-ad})}$, only give the renormalization factor of the second term in Eq. (\ref{eq:adtorque}), so we ignore these contributions. 
\section{Results}
Substituting $F(<\hat{\bm s}(z,t)>)=F^{(\mathrm{ad})}+F^{(\mathrm{non-ad})}$ and $\tau(<\hat{\bm s}(z,t)>)=\tau^{(\mathrm{ad})}$ into Eqs. (\ref{eq:eom2}) and (\ref{eq:eom3}), we obtain the equation of motion of the CSL as
\begin{align}
-\mathcal{K} \left( S+\frac{s}{2} \right) \dot{Z}(t)+\alpha S\dot{\xi}_0(t)+2\epsilon_0^{(\theta)}\xi_0(t) +\frac{\mathcal{K}}{2}\frac{j_s}{e}&=0, \label{eq:eoffinal1} \\
\alpha\mathcal{M}S\dot{Z}(t)+\mathcal{K} \left( S+\frac{s}{2} \right) \dot{\xi}_0(t)-F^{(\mathrm{non-ad})} &=0. \label{eq:eoffinal2}
\end{align}
By eliminating $\dot{Z}(t)$ from Eqs. (\ref{eq:eoffinal1}) and (\ref{eq:eoffinal2}), and imposing the boundary condition $\xi_0(0)=0$, we obtain
\begin{align}
\xi_0(t)&=\xi^*(1-e^{-\lambda t} ),
\end{align}
where
\begin{align}
\lambda=&\frac{2\epsilon_0^{(\theta)}\alpha S}{\left(S+\frac{s}{2}\right)^2+(\alpha S)^2} ,\\
\xi^*=&-\frac{\mathcal{K}}{4\epsilon_0}\frac{j_s}{e} +\frac{\left(S+\frac{s}{2}\right)F^{(\mathrm{non-ad})} }{2\epsilon_0\alpha \mathcal{K}S}. 
\end{align}
Then the velocity of the CSL becomes
\begin{align}
\dot{Z}(t)&=V^*+V_0 e^{-\lambda t},
\end{align}
with
\begin{align}
V_0=&-\frac{\frac{j_s}{2e}\left(S+\frac{s}{2}\right)}{\alpha^2S^2+\left(S+\frac{s}{2}\right)^2}-\frac{\left(S+\frac{s}{2}\right)^2}{\alpha\mathcal{M}S}\frac{F^{(\mathrm{non-ad})}}{(\alpha S)^2+\left(S+\frac{s}{2}\right)^2},\\
V^*=&\frac{1}{\alpha\mathcal{M}S}F^{(\mathrm{non-ad})}. \label{Vstar}
\end{align}

After the relaxation time $\frac{1}{\lambda}$, the velocity of the CSL becomes the terminal velocity $V^*$. It is apparent that $F^{(\mathrm{non-ad})}$ causes a qualitative change in the dynamics; in the adiabatic limit, the terminal velocity is zero because $F^{(\mathrm{non-ad})}=0$. This is different from the domain wall case, where a domain wall can move above a critical current even in the adiabatic limit.\cite{Tatara2004,Tatara2007} The direction of motion is opposite to the current, in other words, in the same direction as the carrier flow. We also point out that the CSL is tilted out of the easy plane by $\delta\theta=u_0(z-Z(t))\xi^*$ under an electric field.\par
\begin{figure}[]
\begin{center}
\includegraphics[width=\hsize]{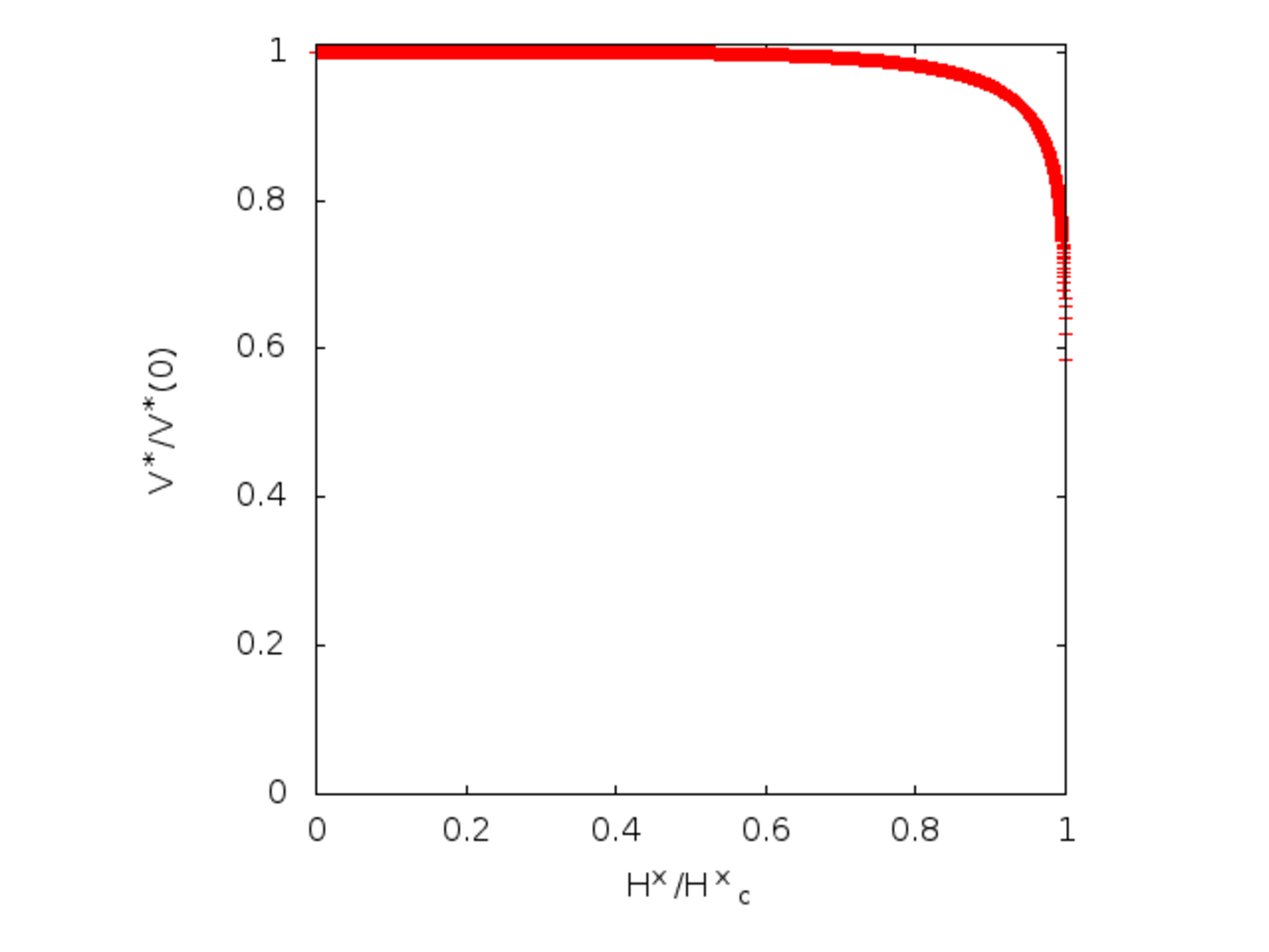}
\end{center}
\caption{Magnetic field dependence of the terminal velocity of the CSL: $V^*$ decreases when the magnetic field increases. Finally, $V^*$ becomes 0. A steep curve arises near $H^x=H^x_c$. The x-axis and y-axis are normalized by $H^x_c$ and $V^*(0)$, respectively.}
\label{fig:V}
\end{figure}

We show the magnetic field ($H^x$) dependence of $V^*$ in Fig. \ref{fig:V}. The terminal velocity of the CSL decreases when the magnetic field increases. It becomes zero when the magnetic field reaches the critical field $H_c^x$, where the system becomes the FFM state. It is natural that the velocity becomes low when the density of solitons decreases because the torque is generated by the spatial modulation of the spin configuration. \par
As already mentioned, we have used an approximation that is valid in the weak-magnetic-field region. Therefore, when $H^x$ is close to $H_c^x$, our model is not exact. However, in the FFM state, it is natural to expect that no motion occurs under an electric field. Thus, it appears that the tendency shown in Fig. \ref{fig:V} near $H_c^x$ is more or less correct. 
To estimate $V^*$ at $H^x=0$, we use typical values $\frac{J_{\mathrm{sd}}}{\epsilon_F}=10^{-1},\frac{ne^2\tau E}{m_e}=10^{8}[\mathrm{A/m}^2],a_0=10^{-10}[\mathrm{m}],k_F=10^{10}[\mathrm{m}^{-1}],D/J=10^{-1},\alpha=10^{-2},n=10^{29}[\mathrm{m}^{-3}]$. In this case, we obtain $V^*(0)\simeq$0.1[m/s], which would be a reasonable value in experiments. \par
Finally, we mention the relation between the terminal velocity and the magnetoresistance. The resistivity $\rho_s$ due to the spin structure is given by \cite{Tatara2008}
\begin{align}
\rho_s=\frac{4\pi J_{\mathrm{sd}}}{e^2n^2}\frac{1}{V}\sum_{\bm k,\bm q,\sigma}|A^\sigma_z(\bm q)|^2\delta(\epsilon_{\bm k\pm})\delta (\epsilon_{\bm k +q-\sigma}-\epsilon_{\bm k\sigma}). \label{eq:rho}
\end{align}
Therefore, we can see
\begin{align}
F^{(\mathrm{non-ad})}=\frac{e^3 E \tau}{m}\rho_s n^2=en\rho_s j.
\end{align}
In other words, $F^{(\mathrm{non-ad})}$ is the reaction of the momentum transfer of conduction electrons which causes the resistance. This is consistent with a previous experiment\cite{Togawa2013} that showed a negative MR in proportion to the CSL density. The $H^x$ dependence in Fig. \ref{fig:V} is not exactly the same as that of the CSL density. To compare with experimental results more closely, we need to consider other contributions to the resistivity originating from the mechanisms other than the present mechanism. This remains as a future problem.\par
Before we finish this section, we mention the pinning effect. In this paper, we ignore the possible effects of pinning. Experimentally, it has been reported \cite{Togawa2012} that the CSL formed in  $\mathrm{CrNb}_3\mathrm{S}_6$ exhibits reasonably robust coherence over macroscopic scales. Therefore, we expect that once the CSL begins to move as a coherent heavy object, a microscopic pinning mechanism may be irrelevant. For example, in the case of a ferromagnetic domain wall, the Barkhausen effect of the magnetization is caused by irreversible magnetic domain wall motion by it breaking away from pinning sites. However, experiments by Tsuruta et al. \cite{Tsuruta2015} clearly indicate that there is no Barkhausen effect for the CSL formed in  $\mathrm{CrNb}_3\mathrm{S}_6$. This fact also suggests the irrelevance of pinning effects. However, at this stage there has been not theoretical study on possible pinning effects to determine their relevance or irrelevance. Thus, we will keep this issue beyond the scope of the present paper.

\section{Conclusion}
In this paper, we extended the theory of CSL motion to a finite magnetic field. We pointed out that the torque from the conduction electrons changes as a function of the external magnetic field. As a result, the terminal velocity of the CSL decreases when the magnetic field increases. 
One of the most important features of the CSL is that we can control the responses by an external magnetic field.
We hope that our tunable local spin dynamics will open a new door in the research of spintronics.

\section*{Acknowledgements}
We thank M. Hayashi, T. Mizoguchi, Y. Onose, A. S. Ovchinnikov, I. Proskurin, and G. Tatara for fruitful discussion.
This work was supported by the CResCent(Chirality Research Center) in Hiroshima University. K. T. was supported by the Japan Society for the Promotion of Science through the Program for Leading Graduate Schools (MERIT).

\end{document}